\documentclass[twocolumn,english,aps,superscriptaddress,floatfix]{revtex4}
\usepackage[T1]{fontenc}
\usepackage[latin9]{inputenc}
\usepackage{color}
\usepackage{bm}
\usepackage{xcolor}
\usepackage{amsmath}
\usepackage{amssymb}
\usepackage{graphicx}
\usepackage{natbib}
\usepackage{braket}
\usepackage{psfrag}
\usepackage{tikz}
\usepackage[normalem]{ulem}
\usepackage{qcircuit}
\usepackage{ulem}

\allowdisplaybreaks

  {{\nopagebreak\hspace*{\fill}$\Box$\par\vspace{12pt}}}

\begin{document}

\title{Efficient State Preparation for Quantum Amplitude Estimation}

\author{Almudena Carrera Vazquez}
\affiliation{%
IBM Research -- Zurich
}%
\affiliation{%
ETH Zurich
}%
\author{Stefan Woerner}
\email{wor@zurich.ibm.com}
\affiliation{%
IBM Research -- Zurich
}%


\date{\today}

\begin{abstract}
Quantum Amplitude Estimation (QAE) can achieve a quadratic speed-up for applications classically solved by Monte Carlo simulation.
A key requirement to realize this advantage is efficient state preparation.
If state preparation is too expensive, it can diminish the quantum advantage.
Preparing arbitrary quantum states has exponential complexity with respect to the number of qubits, thus, is not applicable.
Currently known efficient techniques require problems based on log-concave probability distributions, involve learning an unknown distribution from empirical data, or fully rely on quantum arithmetic.
In this paper, we introduce an approach to simplify state preparation, together with a circuit optimization technique, both of which can help reduce the circuit complexity for QAE state preparation significantly.
We demonstrate the introduced techniques for a numerical integration example on real quantum hardware, as well as for option pricing under the Heston model, i.e., based on a stochastic volatility process, using simulation.
\end{abstract}

\maketitle

\section{\label{sec:introduction} Introduction}

Quantum Amplitude Estimation (QAE) is a quantum algorithm that can achieve a quadratic speed-up over classical Monte Carlo simulation \cite{Brassard2000, Montanaro2017, Suzuki2019, Aaronson2019, Grinko2019}.
It has many possible applications, such as option pricing or risk analysis \cite{Rebentrost2018, Woerner2019, Egger2019, Stamatopoulos2019}, or numerical integration \cite{Abrams1999}.

A key requirement to apply QAE is to be able to efficiently load the problem of interest.
In other words, we need a quantum circuit with a depth polynomial in the number of qubits that prepares a quantum state corresponding to the problem we like to solve.
This can be achieved by using quantum arithmetic, which, although polynomial in the number of qubits, usually requires a significant overhead in terms of gates and ancilla qubits \cite{Haner2018}.
Another approach is to directly prepare a quantum state that corresponds to a probability distribution.
However, preparing generic quantum states requires an exponential number of gates \cite{shende2006, Plesch2010}, and is thus not applicable here since it would diminish the quantum advantage.
Efficient approaches either require the distribution to be log-concave \cite{Grover2002}, leverage quantum machine learning techniques to train a quantum operator to approximate an unknown underlying distribution given empirical data \cite{Zoufal2019}, or approximate smooth, differentiable functions using piecewise polynomial approximations and matrix product states \cite{holmes2020efficient}.

In the present paper, we improve the known techniques and show how to simplify multiplication and addition of functions on the amplitude level, i.e., reducing part of the overhead  introduced by quantum arithmetic.
Furthermore, we introduce a circuit optimization technique that helps to asymptotically halve the circuit depth for QAE.
We discuss QAE from a numerical integration point of view, which allows us to use standard error estimates from numerical integration and to study how the model error reduces with respect to the number of qubits used.
Finally, we demonstrate the introduced techniques, first by using a numerical integration experiment on a real quantum device, and second, by showing how to price a European call option under the Heston model, i.e., considering a stochastic volatility process for the price of the underlying asset.

The remainder of this paper is structured as follows.
Sec.~\ref{sec:qae} defines QAE and the corresponding state preparation problem.
Sec.~\ref{sec:state_preparation} introduces a more efficient state preparation scheme and extends it to stochastic processes.
Sec.~\ref{sec:spin_echo} shows how the resulting quantum circuits can be significantly simplified.
In Sec.~\ref{sec:error_analysis} we analyze the error resulting from approximating a continuous function using a finite number of qubits, i.e., grid points. We link it to basic numerical integration and show how to reduce this error while keeping the number of qubits constant.
Sec.~\ref{sec:results} illustrates our results using both simulation and quantum hardware, and Sec.~\ref{sec:conclusions} concludes the paper.

\section{\label{sec:qae} Quantum Amplitude Estimation}

Suppose an operator $\mathcal{A}$ acting on $n+1$ qubits as
\begin{eqnarray}
\mathcal{A}\ket{0}_n\ket{0} = \sqrt{1-a}\ket{\psi_0}_n\ket{0} + \sqrt{a}\ket{\psi_1}_n\ket{1}, \label{eq:A}
\end{eqnarray}
where $\ket{\psi_0}, \ket{\psi_1}$ are normalized quantum states, and $a \in [0, 1]$ is the probability of measuring the last qubit in state $\ket{1}$.
Following on from \cite{Brassard2000}, we will call states with $\ket{1}$ in the last qubit \emph{good states}, and the others \emph{bad states}.

Accordingly, QAE is a quantum algorithm that allows to estimate $a$ with a quadratic speed-up over classical Monte Carlo simulation. The algorithm repeatedly applies the operator $\mathcal{Q} = \mathcal{A}\mathcal{S}_0\mathcal{A}^{\dagger}\mathcal{S}_{\psi_0}$ to $\mathcal{A}\ket{0}_{n+1}$, where $\mathcal{S}_0 = \mathbb{I}_{n+1} - 2\ket{0}\bra{0}_{n+1}$  and $\mathcal{S}_{\psi_0} = \mathbb{I}_{n+1} - 2\ket{\psi_0}\ket{0}\bra{\psi_0}\bra{0}$ are reflections and $\mathbb{I}_{n+1}$ is the identity operator on $n+1$ qubits.
The resulting error scales as $\mathcal{O}(1/M)$, where $M$ denotes the number of (quantum) samples, i.e., applications of $\mathcal{Q}$.
In contrast, the error resulting from Monte Carlo simulation scales as $\mathcal{O}(1/\sqrt{M})$ for $M$ (classical) samples \cite{Brassard2000,Glasserman2003}.

The canonical form of QAE requires controlled applications of $\mathcal{Q}$ within quantum phase estimation \cite{Brassard2000}.
Recently, different, simpler variants have been proposed, which only require to run $\mathcal{Q}^k\mathcal{A}\ket{0}_{n+1}$ for different powers $k$, such as the \emph{Maximum Likelihood Amplitude Estimation} (MLAE) \cite{Suzuki2019} or the \emph{Iterative Quantum Amplitude Estimation} (IQAE) \cite{Grinko2019}.
In the remainder of this paper we will consider MLAE for demonstrations, although our results are applicable to every variant of QAE.
MLAE does not provide a theoretical guarantee on the result, however, it performs well in practice and is well suited as benchmark, since it allows freedom on the choice of $k$.
For a discussion and comparison of different variants of QAE, we refer to \cite{Grinko2019}.

Due to the definition of $\mathcal{A}$ in (\ref{eq:A}), $\mathcal{S}_{\psi_0}$ can easily be constructed by only considering the ancilla qubit.
Note that the original formulation of QAE does not require \emph{good states} and \emph{bad states} to be flagged by an ancilla qubit but is more generic \cite{Brassard2000}.
In general, it may be a bit more complex to implement the reflection $\mathcal{S}_{\psi_0}$, but the underlying theory holds as well.
We will leverage this fact later, when we flag \emph{good states} and \emph{bad states} using an encoding of multiple ancilla qubits.

The canonical QAE is based on Quantum Phase Estimation (QPE), which introduces an overhead in terms of the number of required qubits and circuit depth.
However, variants of QAE that can achieve a quadratic speedup without QPE \cite{Suzuki2019, Aaronson2019, Grinko2019} have recently been proposed.
In order to apply these algorithms to a more generic problem than (1), we need to adjust $S_{\psi_0}$ to identify \emph{good states} and \emph{bad states}, and to be able to decide from a single measurement of all qubits whether we observed part of a \emph{good state} or not.
The latter can be achieved by using an encoding of multiple ancilla qubits.

A common way to construct $\mathcal{A}$ is to first load a probability distribution and then apply an objective function, as outlined in the following.
Assume $n$ qubits, a random variable $X$ defined by the possible values $x_i = a \cdot i + b$, $a, b \in \mathbb{R}$, with their corresponding probabilities $p_i \in [0, 1]$, $i=0, ..., 2^n-1$, and a quantum operator $\mathcal{U}$ acting as
\begin{eqnarray}
\mathcal{U} \ket{0}_n = \sum_{i=0}^{2^n-1} \sqrt{p_i}\ket{i}_n. \label{eq:load_prob_U}
\end{eqnarray}
Furthermore, suppose an objective function $g: \mathbb{R} \rightarrow [0, 1]$, one additional qubit in state $\ket{0}$, and a corresponding quantum operator $\mathcal{G}$ defined by
\begin{eqnarray}
\mathcal{G}: \ket{i}_n\ket{0} \mapsto \ket{i}_n \left(\sqrt{1 - g(x_i)}\ket{0} + \sqrt{g(x_i)}\ket{1} \right).
\end{eqnarray}
If we set $\mathcal{A} = \mathcal{G} ( \mathcal{U} \otimes \mathbb{I} )$ and apply it to $\ket{0}_{n+1}$, then the probability of measuring $\ket{1}$ in the last qubit is given by
\begin{eqnarray}
\sum_{i=0}^{2^n-1} p_i g(x_i),
\end{eqnarray}
which is equal to the expected value $\mathbb{E}[g(X)]$.

Thus, if we can construct $\mathcal{U}$ and $\mathcal{G}$ efficiently, we can achieve a quadratic speed-up to estimate $\mathbb{E}[g(X)]$ by using QAE for $\mathcal{A}$.
Efficient ways to approximate $\mathcal{G}$ for polynomial $g$ are discussed in \cite{Woerner2019, Stamatopoulos2019}.
For more general functions, $\mathcal{G}$ can be constructed using quantum arithmetic, by first computing $\sin^{-1}(\sqrt{g(x_i)})$ into an ancilla qubit register and then using controlled $Y$-rotations to prepare the amplitude of the ancilla qubit.

However, as discussed in Sec.~\ref{sec:introduction}, constructing $\mathcal{U}$ is more challenging, and in general requires an exponential number of gates.
In the following, we show how an alternative approach that allows to efficiently construct $\mathcal{A}$ whenever the probabilities $p_i$ and $g$ are given by efficiently computable functions.
We show how this extends to multivariate problems as well as stochastic processes.

\section{\label{sec:state_preparation} Efficient State Preparation}

Let us first assume the simple case where $p_i = 1/2^n$, i.e., $X$ follows a uniform distribution.
This is easy to prepare by applying Hadamard gates to all $n$ state qubits.
Then, applying $\mathcal{G}$, defined as before, leads to the state
\begin{eqnarray}
\frac{1}{\sqrt{2^n}} \sum_{i=0}^{2^n-1} \ket{i}_n \left( \sqrt{1 - g(x_i)}\ket{0} + \sqrt{g(x_i)}\ket{1} \right),
\label{eq:uniform_riemann_state}
\end{eqnarray}
with the probability of measuring $\ket{1}$ in the ancilla qubit being equal to
\begin{eqnarray}
\frac{1}{2^n} \sum_{i=0}^{2^n-1} g(x_i),
\end{eqnarray}
as proposed in \cite{Montanaro2017}.
This can be interpreted as a \emph{left Riemann sum} \cite{DAVIS198451}, i.e., a $2^n$-point approximation of the integral $\int_{x=0}^1 g(x) dx$, assuming $x_i = i/2^n$.

Next, suppose a random variable $X$ with a corresponding probability density function (PDF) $f: \mathbb{R} \rightarrow \mathbb{R}_{\geq 0}$.
We can replace $g$ in (\ref{eq:uniform_riemann_state}) by the product of $f$ and $g$, which allows us to approximate the expectation value $\mathbb{E}_f[g(X)]$.
Depending on $f$, we may need to normalize the problem such that $f(x)g(x) \in [0, 1]$ for all $x$.
In the following, we introduce an alternative approach to estimate $\mathbb{E}_f[g(X)]$, which can easily be extended to stochastic processes.

As just introduced, suppose a random variable $X$ and the corresponding (normalized) PDF $f: \mathbb{R} \rightarrow [0, 1]$.
Furthermore, suppose $n$ state qubits and two ancilla qubits.
Following the approaches outlined in Sec.~\ref{sec:qae}, we can prepare operators $\mathcal{F}$ and $\mathcal{G}$ such that
\begin{equation}
\begin{split}
\mathcal{F}: &\ket{i}_n\ket{0}\ket{j} \mapsto \\
&\ket{i}_n\left( \sqrt{1-f(x_i)}\ket{0} + \sqrt{f(x_i)}\ket{1} \right) \ket{j},
\end{split}\label{eq:F_operator}
\end{equation}
and
\begin{equation}
\begin{split}
\mathcal{G}: &\ket{i}_n\ket{j}\ket{0} \mapsto \\
&\ket{i}_n\ket{j}\left( \sqrt{1-g(x_i)}\ket{0} + \sqrt{g(x_i)}\ket{1} \right),
\end{split}
\end{equation}
where $\mathcal{F}$ prepares the first ancilla qubit and $\mathcal{G}$ prepares  the second one.
We now apply Hadamard gates to the first $n$ qubits of $\ket{0}_n \ket{00}$ followed by $\mathcal{F}$ and $\mathcal{G}$, which leads to the state
\begin{eqnarray}
\begin{aligned}
&\frac{1}{\sqrt{2^n}} \sum_{i=0}^{2^n-1} \ket{i}_n \sqrt{1-f(x_i)} \sqrt{1 - g(x_i)} \ket{00} \\
+&\frac{1}{\sqrt{2^n}} \sum_{i=0}^{2^n-1} \ket{i}_n \sqrt{1-f(x_i)} \sqrt{g(x_i)} \ket{01} \\
+&\frac{1}{\sqrt{2^n}} \sum_{i=0}^{2^n-1} \ket{i}_n \sqrt{f(x_i)} \sqrt{1 - g(x_i)} \ket{10} \\
+&\frac{1}{\sqrt{2^n}} \sum_{i=0}^{2^n-1} \ket{i}_n \sqrt{f(x_i)} \sqrt{g(x_i)} \ket{11}.
\end{aligned}
\end{eqnarray}
Subsequently, the probability of measuring $\ket{11}$ for the last two qubits is given by
\begin{eqnarray}
\frac{1}{2^n} \sum_{i=0}^{2^n-1} f(x_i)g(x_i),
\end{eqnarray}
which is again equal to a Riemann sum and approximates the expected value $\mathbb{E}_f[g(X)]$.
In this case, we define $\mathcal{S}_{\psi_0} = (\mathbb{I}_{n+2} - 2\ket{\psi_1}\ket{11}\bra{\psi_1}\bra{11})$ to construct $\mathcal{Q}$, i.e., \emph{good states} and \emph{bad states} are identified by the two ancilla qubits and we do not have to explicitly multiply $f$ and $g$ using quantum arithmetic.
Note that we deviate from the usual definition here, i.e., we multiply the \emph{good states} ($\ket{\psi_1}$) by $-1$, instead of the \emph{bad states} ($\ket{\psi_0}$).
However, this only implies a difference in the global phase and can be ignored in the following.

Similar to multiplication, we can also construct operators to realize addition of functions.
Suppose the operators $\mathcal{G}$ and $\mathcal{H}$, corresponding to functions $g$ and $h$, that not only share the control qubits but also the target qubit.
Furthermore, let us add an ancilla qubit in state $\ket{0}$ and consider the initial state $\ket{i}_n\ket{0}\ket{0}$.
Applying a Hadamard gate to the added ancilla and then the two (controlled) operators
\begin{align}
\mathcal{G} &\otimes \ket{0}\bra{0} + \mathbb{I}_{n+1}\ket{1}\bra{1}, \\
\mathcal{H} &\otimes \ket{1}\bra{1} + \mathbb{I}_{n+1}\ket{0}\bra{0},
\end{align}
leads to the state
\begin{eqnarray}
\begin{aligned}
&\frac{1}{\sqrt{2}}
\ket{i}_n \left(\sqrt{1-g(x_i)}\ket{0} + \sqrt{g(x_i)}\ket{1}\right)\ket{0}\\
+&\frac{1}{\sqrt{2}}
\ket{i}_n \left(\sqrt{1-h(x_i)}\ket{0} + \sqrt{h(x_i)}\ket{1}\right)\ket{1}.
\end{aligned}
\end{eqnarray}
Applying another Hadamard gate to the ancilla qubit transforms the state into
\begin{eqnarray}
\begin{aligned}
&\frac{1}{2} \left(\sqrt{1-g(x_i)} + \sqrt{1-h(x_i)}\right)
\ket{i}_n\ket{0}\ket{0}\\
+&\frac{1}{2} \left(\sqrt{1-g(x_i)} - \sqrt{1-h(x_i)}\right)
\ket{i}_n\ket{0}\ket{1}\\
+&\frac{1}{2} \left(\sqrt{g(x_i)} + \sqrt{h(x_i)}\right)
\ket{i}_n\ket{1}\ket{0}\\
+&\frac{1}{2} \left(\sqrt{g(x_i)} - \sqrt{h(x_i)}\right)
\ket{i}_n\ket{1}\ket{1}.
\end{aligned}
\end{eqnarray}
If we now define the \emph{good states} again as those with $\ket{1}$ in the target qubit, i.e., the second to last qubit, then the probability of measuring these states equals $(g(x_i) + h(x_i))/2$, i.e., we can add the functions $g$ and $h$.
This construction is closely related to the \emph{Linear Combination of Unitaries} (LCU) \cite{Berry_2015}.
However, LCU requires the ancilla qubit to be measured in a particular state to get the desired (non-unitary) operation.
In the present context, this is not necessary, since we can adjust $\mathcal{S}_{\psi_0}$ accordingly within QAE.
Note that the addition introduces a factor of $1/2$ that we need to take into account by multiplying the resulting estimate by a factor of $2$.
This also increases the estimation error accordingly, which means that the error increases exponentially with the number of additions, which might limit the number of settings where this leads to a favorable scaling.
Note that the expected value is linear, i.e., we may also realize a sum by estimating the terms individually.

\begin{figure}[ht]
\centering
\includegraphics[width=0.49\textwidth]{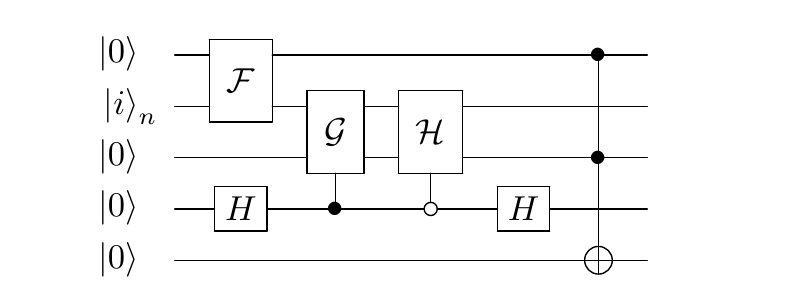}
\caption{
Addition and multiplication of amplitudes for QAE: Suppose functions $f, g, h$ and the corresponding operators $\mathcal{F}, \mathcal{G}, \mathcal{H}$, as defined in (\ref{eq:F_operator}), all sharing the control qubits $\ket{i}_n$, but with the difference that $\mathcal{F}$ targets the qubit above $\ket{i}_n$ and the other two operators the qubit below $\ket{i}_n$.
The illustrated circuit then prepares a state with the probability of measuring the bottom qubit in state $\ket{1}$ equal to $f(x_i)(g(x_i) + h(x_i))$.
The Toffoli gate is necessary in case another function should be added, otherwise, adjusting $\mathcal{S}_{\psi_0}$ is sufficient.
}
\label{fig:fgh_add_mult}
\end{figure}

Exploiting the presented approach, we can construct arbitrary combinations of additions and multiplications of functions for which we have oracles of the form given for $\mathcal{F}$, $\mathcal{G}$, $\mathcal{H}$.
In case an addition should take place after one or more multiplications, i.e., in situations where the \emph{good states} are flagged by multiple ancilla qubits being in state $\ket{1}$, it might be necessary to add an additional ancilla qubit, and apply a multi-controlled NOT gate to reduce back to a single qubit flagging the \emph{good states}.
A corresponding circuit is illustrated in Figure \ref{fig:fgh_add_mult}.

So far, we focused on univariate problems.
It is straightforward to extend this approach to multivariate problems as well, e.g., by representing the dimensions by separate quantum registers, possibly with a different number of qubits each.
The required number of qubits will scale linearly in the dimension, unlike classical numerical integration schemes where the dependence is usually exponential -- except for Monte Carlo simulation.
In the following, we will show how this approach can be extended to load stochastic processes.

Suppose a stochastic process given by an initial probability distribution $f_0(x^0)$ and the transition probabilities $f_t(x^t \mid x^{t-1})$, i.e., the probability to reach state $x^t$ at time $t$, given the state history $x^{t-1}$.
Here, to simplify the notation, we assume Markov processes, i.e., $f_t$ only depends on $x^t$ and $x^{t-1}$, not the full history of the process.
However, it is straightforward to extend our approach to generic (discretized) stochastic processes.
Furthermore, suppose $n_t$ qubits to represent the state $x^t$ at time $t$, for $t=0, \dots T$, add $T+1$ ancilla qubits, and define $n = n_0 + \cdots + n_T$.
Then, as in (\ref{eq:F_operator}), we can construct an operator $\mathcal{F}_0$ corresponding to $f_0$ that prepares the first ancilla qubit.
In addition, we can construct operators
\begin{eqnarray}
\begin{aligned}
\mathcal{F}_t:
&\ket{i}_{n_{t-1}}\ket{j}_{n_{t}}\ket{0} \mapsto \\
&\ket{i}_{n_{t-1}}\ket{j}_{n_{t}} \\
&
\left(
\sqrt{1 - f_t\left(x_j^t \mid x_i^{t-1}\right)}\ket{0} + \sqrt{f_t\left(x_j^t \mid x_i^{t-1}\right)}\ket{1}
\right).
\end{aligned}
\end{eqnarray}
If we first apply Hadamard gates to all state qubits and then apply $\mathcal{F}_0, \ldots, \mathcal{F}_T$ to the corresponding qubit registers and ancilla qubits, we can construct the $(n+T+1)$-qubit state
\begin{eqnarray}
\begin{aligned}
\frac{1}{\sqrt{2^n}} &\sum_{i_0=0}^{2^{n_0-1}} \cdots \sum_{i_T=0}^{2^{n_T-1}} \ket{i_0}_{n_0}\cdots\ket{i_T}_{n_T} \\
&\left[ \ldots + \sqrt{f_0\left(x^0_{i_0}\right) \prod_{t=1}^T f_t\left( x_{i_t}^{t} \mid x_{i_{t-1}}^{t-1} \right)}
\ket{1 \ldots 1}_{T+1} \right],
\end{aligned}
\end{eqnarray}
where we drop the terms without $\ket{1 \ldots 1}_{T+1}$ in the ancilla qubits.

Given an objective function $g: \mathbb{R}^{T+1} \rightarrow [0, 1]$ and the corresponding operator $\mathcal{G}$, we can add another ancilla qubit and apply $\mathcal{G}$.
The resulting probability of measuring $\ket{1 \ldots 1}_{T+2}$ in all ancilla qubits is then given by
\begin{eqnarray}
\begin{aligned}
\frac{1}{2^{n}} \sum_{i_0, \ldots, i_T}
& f_0\left(x^0_{i_0}\right) \prod_{t=1}^T f_t\left( x_{i_t}^{t} \mid x_{i_{t-1}}^{t-1} \right)
g\left(x^0_{i_0}, \ldots, x^T_{i_T}\right),
\end{aligned}
\end{eqnarray}
which approximates the expectation value $\mathbb{E}[g(X)]$ where $X = (x^0, \ldots x^T)$ represents the possible paths of the (discretized) stochastic process defined by the $f_t$.

Note that the complexity of $S_{\psi_0}$ essentially equals the complexity of a multi-controlled NOT gate applied to the target qubits.
Thus, its gate complexity is always less than the complexity of $S_0$, which corresponds to a multi-controlled NOT gate that is controlled by all but one qubit (excluding work qubits that are used and then cleaned within the construction of $\mathcal{A}$, e.g., to realize quantum arithmetic). 
A detailed analysis of the complexity of implementing $S_0$ and $S_{\psi_0}$ can be found in \cite{Egger2019}.

In principle, stochastic processes could also be loaded using a similar approach to (\ref{eq:load_prob_U}).
However, that would require the construction of operators
\begin{align}
\mathcal{U}_t: 
&\ket{x_{t-1}}_{n_{t-1}}\ket{0}_{n_t} \mapsto \nonumber \\
&\ket{x_{t-1}}_{n_{t-1}}\sum_{i_t=0}^{2^{n_t}-1} \sqrt{f_t(x_{i_t \mid x_{t-1}})} \ket{x_{i_t}}_{n_t},
\end{align}
which, like $\mathcal{U}$ in (\ref{eq:load_prob_U}), cannot be done efficiently for generic processes.

\section{\label{sec:spin_echo} Spin-Echo Circuit Optimization}

Within this section we present a technique to optimize the circuits resulting from the construction introduced in Sec.~\ref{sec:state_preparation}.
More precisely, we leverage an effect that is also known as \emph{Spin-Echo} \cite{Das1954}.
In some cases, this can help to significantly reduce the gates required to construct $\mathcal{Q}^k \mathcal{A} \ket{0}_n\ket{0}$.

Suppose circuits of the form
\begin{eqnarray}
R_U(\theta) V R_U(-\theta),
\end{eqnarray}
where $U, V \in \{X, Y, Z\}$, i.e., $R_U$ denotes a single-qubit Pauli rotation and $V$ a single-qubit Pauli gate.
In case $U = V$, the gates commute and the circuit equals $V$.
In all other cases, it can easily be seen that
\begin{eqnarray} \label{eq:single_qubit_spin_echo}
R_U(\theta) V R_U(-\theta) = R_U(2\theta) V,
\end{eqnarray}
which is called the \emph{Spin-Echo}.
Note that the right-hand side could also be written as $V R_U(-2\theta)$.

For a given function $f: \mathbb{R} \rightarrow \mathbb{R}$ let us define an operator $R_U^f$ on $n+1$ qubits as
\begin{eqnarray}
R_U^f: \ket{x}_n\ket{\phi} \mapsto \ket{x}_nR_U(f(x))\ket{\phi},
\end{eqnarray}
for an arbitrary single-qubit state $\ket{\phi}$ and $x \in \{0, \ldots 2^n-1\}$.
If we replace $R_U$ in (\ref{eq:single_qubit_spin_echo}) by $R_U^f$, the same identity holds and we have
\begin{eqnarray} \label{eq:spin_echo}
R_U^f (\mathbb{I}_n \otimes V) R_U^{(-f)} = R_U^{(2f)} (\mathbb{I}_n \otimes V),
\end{eqnarray}
where $V$ is applied to the last qubit only.

\begin{figure}[ht]
\centering
\includegraphics[width=0.49\textwidth]{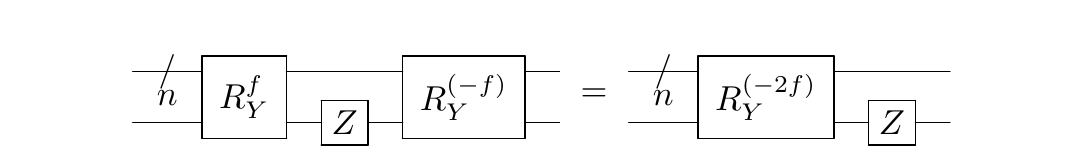}
\caption{
Spin-Echo circuit optimization for $U=Y$ and $V=Z$.
}
\label{fig:spin_echo_circuit}
\end{figure}

Suppose now that we want to use QAE to estimate the integral of $f$.
We can set $U = Y$ and construct $\mathcal{A} = R_Y^f  (H^{\otimes n} \otimes \mathbb{I})$.
Furthermore, we set $V = Z$, which then corresponds to $\mathcal{S}_{\psi_0}$ in the definition of $\mathcal{Q}$.
When constructing $\mathcal{Q}^k \mathcal{A} \ket{0}_n\ket{0}$, we repeatedly have the pattern
$\mathcal{A}^{\dagger} \mathcal{S}_{\psi_0} \mathcal{A}$, $k$ times in total, which equals
\begin{eqnarray}
(H^{\otimes n} \otimes \mathbb{I}) R_Y^{(-f)} (\mathbb{I}_n \otimes Z) R_Y^f (H^{\otimes n} \otimes \mathbb{I}).
\end{eqnarray}
The Hadamard gates at the beginning and the end are dominated by $R_Y^f$ in terms of circuit complexity and we can ignore them in the following analysis.
Thus, we have
\begin{eqnarray}
R_Y^{(-f)} (\mathbb{I}_n \otimes Z) R_Y^f,
\end{eqnarray}
which, following (\ref{eq:spin_echo}), can be simplified to
\begin{eqnarray}
R_Y^{(-2f)} (\mathbb{I}_n \otimes Z),
\end{eqnarray}
as illustrated in Fig.~\ref{fig:spin_echo_circuit}.
In other words, we can drop the $R_Y^f$ in $\mathcal{A}$ and include its effect in the corresponding part of $\mathcal{A}^{\dagger}$ by doubling the rotation angle.
This means that $\mathcal{Q}^k\mathcal{A}$ can be constructed using only $k+1$ instead of $2k+1$ applications of $\mathcal{A}$ (again ignoring $H^{\otimes n}$), i.e., essentially a reduction by a factor of two for larger $k$.
Since, for larger problems, the complexity of $\mathcal{A}$ will dominate the overall complexity \cite{Egger2019}, this directly translates to a corresponding reduction of the circuit depth for QAE.

This circuit optimization is compatible with the addition and multiplication techniques introduced in Sec.~\ref{sec:spin_echo}.
Suppose, e.g., an operator $R_Y^{f(g+h)}$ corresponding to the circuit in Fig.~\ref{fig:fgh_add_mult}, the Spin-Echo circuit optimization can be applied and allows us to remove half of the applications of $R_Y^{f(g+h)}$ in $\mathcal{Q}^k \mathcal{A}$.

\section{\label{sec:error_analysis} Error Analysis \& Higher Order Schemes}

In this section we consider QAE in the context of numerical integration and analyze the approximation error resulting from discretization using $n$ qubits, denoted by $E_n$.
Using basic insights from numerical quadrature, we show how to reduce the approximation error without increasing the  number of discretization qubits.
The discretization error needs to be added on top of the QAE estimation error, since QAE does not estimate the exact value but only an approximation.
The total resulting error behaves like $E_n + \mathcal{O}(1/M)$.
Understanding all error terms is crucial to balancing the number of qubits used to discretize with the target accuracy set for QAE.

Leveraging different approaches from numerical integration allows trading off classical repetitions of QAE with the resulting estimation error $E_n$.
For more details on numerical integration, we refer to \cite{DAVIS198451}.

First, we consider the univariate problem $\int_{x=0}^1 g(x) dx$, for $g: [0, 1] \rightarrow [0, 1]$.
Suppose we use $n$ qubits to discretize the support of $g$, i.e., we use $2^n$ equally-spaced grid points $x_i = i/2^n$, $i=0, \ldots, 2^n-1$.
Denote the left Riemann sum by $R_n^{\text{left}}$, where we use the number of qubits $n$ as index instead of the number of grid points $2^n$.
For simplicity, we assume throughout this section that $g$ is an analytical function, i.e., continuously differentiable.

For $R_n^{\text{left}}$ we know that the estimation error $E_n^{\text{left}}$, defined as
\begin{eqnarray}
E_n^{\text{left}} = \left| \mathbb{E}[g(X)] - R_n^{\text{left}} \right|,
\end{eqnarray}
is bounded by
\begin{eqnarray}
E_n^{\text{left}} \leq \frac{1}{2} \frac{\max_{x \in [0, 1]} |\partial_x g(x)|}{2^n}.
\end{eqnarray}
In other words, the discretization error decreases exponentially with the number of qubits.
The same holds true if we set $x_i = (i+1)/2^n$, i.e., if we evaluate the \emph{right Riemann sum} $R_n^{\text{right}}$.

If we define $x_i = (i + 1/2)/2^n$ instead, we are evaluating the \emph{Midpoint rule}, which results in $R_n^{\text{mid}}$.
This leads to a better scaling, since the estimation error $E_n^{\text{mid}}$ is bounded by
\begin{eqnarray}
E_n^{\text{mid}} \leq \frac{1}{24} \frac{\max_{x \in [0, 1]} |\partial_x^2g(x)|}{2^{2n}},
\end{eqnarray}
i.e., the error drops quadratically faster than for $R_n^{\text{left}}$ or $R_n^{\text{right}}$, while the algorithm has exactly the same complexity and uses the same number of qubits.

The average of $R_n^{\text{left}}$ and $R_n^{\text{right}}$ leads to the \emph{Trapezoidal rule} as well as the corresponding estimator $R_n^{\text{trapez}}$, whose estimation error can be bounded by double the bound for the Midpoint rule but requires two runs of QAE, one for the left and one for the right Riemann sum.

Taking the weighted average of the Trapezoidal rule and the Midpoint rule $(2 R_n^{\text{mid}} + R_n^{\text{trapez}})/3$ leads to \emph{Simpson's rule}, with the resulting estimator $R_n^{\text{Simpson}}$.
Simpson's rule leads to an even better scaling, since the estimation error $E_n^{\text{Simpson}}$ is bounded by
\begin{eqnarray}
E_n^{\text{Simpson}} \leq \frac{1}{2880} \frac{\max_{x \in [0, 1]} |\partial_x^4g(x)|}{2^{4n}},
\end{eqnarray}
i.e., by running QAE three times, we can significantly improve the estimation error.

Other quadrature rules for numerical integration, e.g., \emph{Romberg's method} / \emph{Richardson extrapolation}, are also possible and lead to even higher orders of convergence, while keeping the number of grid points and qubits constant.
Our approach can also be extended to non-equidistant grids, which allows the use of more advanced quadrature schemes to improve the performance, such as Gaussian quadrature \cite{DAVIS198451}.
However, non-equidistant grids require computing the grid points using quantum arithmetic before evaluating the function $g$, leading to longer circuits than equidistant grids and complicating the performance comparison.
Computing the grid points first would also allow to extend the scheme to approximate integrals over infinite domains.

Left and right Riemann sums as well as the Midpoint rule can be easily extended to multivariate problems. 
For instance, suppose a $d$-dimensional function $g: [0, 1]^{d} \rightarrow [0, 1]$.
The resulting error for the Midpoint rule $E^{\text{mid}}_{n,d}$ is bounded by
\begin{eqnarray}
E^{\text{mid}}_{n,d} &\leq& \frac{1}{24} \sum_{i=1}^d
\frac{\max_{x \in [0, 1]^d} |\partial_{x_i}^2 g(x)|}{2^{2n}},
\end{eqnarray}
assuming $n$ qubits per dimension, and analogously for the left and right Riemann sums.
For all three rules, we can achieve an exponential number of grid-points compared to the number of discretization qubits.
Thus, unlike in the classical setting, they also scale efficiently for high-dimensional integrals.
Extending the other approaches to higher dimensions does not scale as favorably, since, e.g., the number of combinations of left and right Riemann sums for different dimensions to evaluate the Trapezoidal rule increases exponentially in $d$.

\section{\label{sec:results} Results}

Within this section, we demonstrate the developed methodology.
First, we perform a small numerical integration experiment on real quantum hardware, leveraging Spin-Echo circuit optimization.
Second, we show how to load the \emph{Heston model}, a stochastic volatility model from mathematical finance, and use it to price a European call option.
Both test cases are implemented in \emph{Qiskit} \cite{Qiskit}.

\subsection{\label{sec:hardware_results} Quantum Hardware: Numerical Integration}

Suppose the integral
\begin{eqnarray}
\int_{x=0}^y \sin^2(\pi x) dx &=& \frac{2\pi y - \sin(2\pi y)}{4\pi},
\end{eqnarray}
for $y \in [0, 1]$, and define $g(x) = \sin^2(\pi x)$ for further reference.
In the following, we will use the  methodology introduced in this paper to approximate the integral using two and three qubits on real hardware.
This analysis is similar to the simulation study in \cite{Suzuki2019} and we will compare the circuit complexity after Spin-Echo optimization to the numbers reported in \cite{Suzuki2019}.

\begin{figure}[ht]
    \centering
    \includegraphics[width=0.35\textwidth]{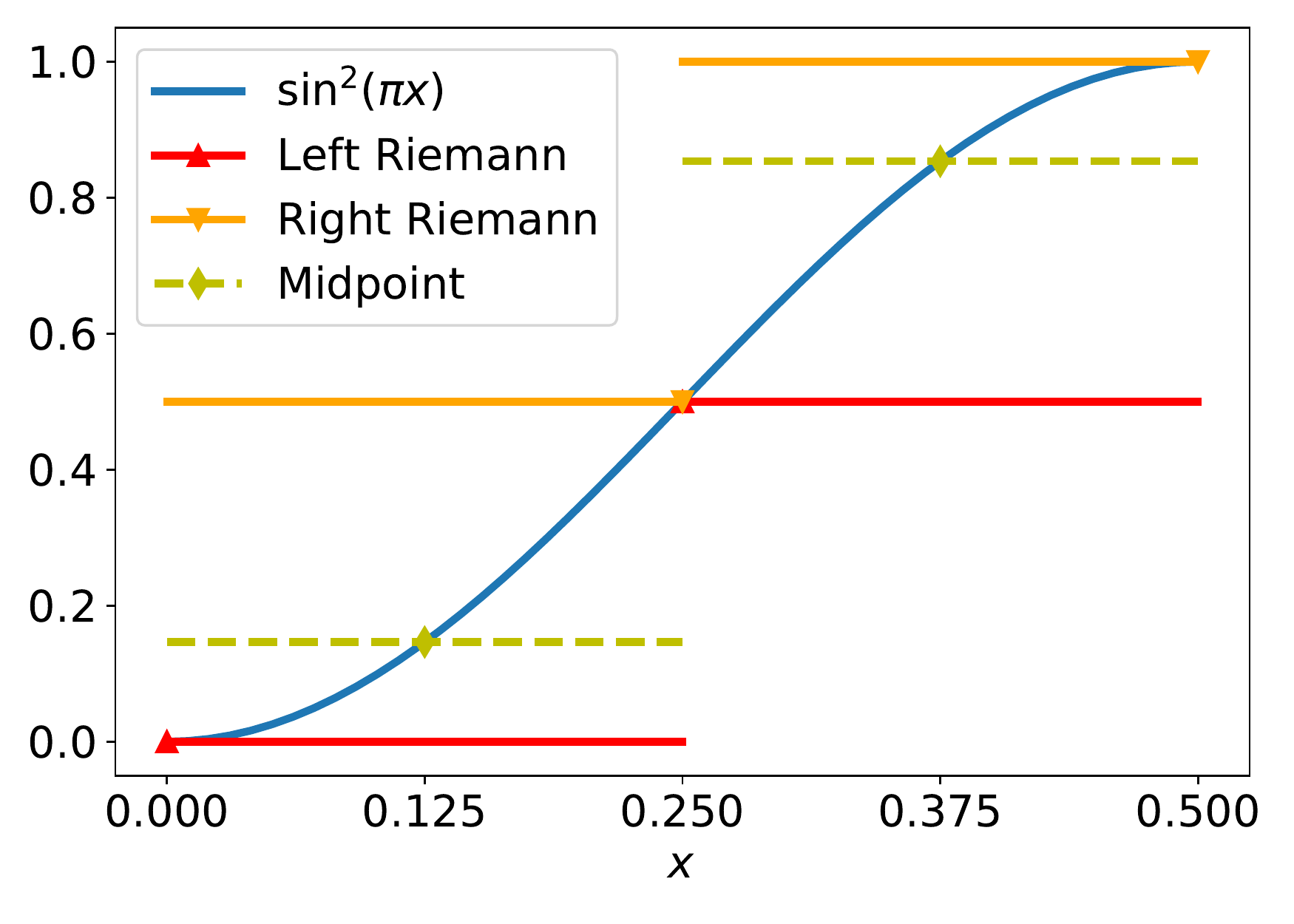}
    \caption{Illustration of the left Riemann sum, right Riemann sum, and Midpoint rule for $y = 1/2$, i.e., to approximate $\int_{x=0}^{1/2} \sin^2(\pi x) dx$.}
    \label{fig:sin2_plot}
\end{figure}

For a given $y \in [0,1]$, we discretize the interval $[0, y]$ to approximate the integral.
We will use one and two qubits for the discretization, i.e., two and four discretization points, and an additional qubit to represent the function $g$.
We define the grid points $x_i$, $i=0, \ldots, 2^n-1$, for $n$ discretization qubits, according to the left Riemann sum, the right Riemann sum, and the Midpoint rule, respectively.
Having the results for all three quadrature rules also allows us to evaluate the corresponding values for Simpson's rule.
Fig.~\ref{fig:sin2_plot} illustrates the considered problem for $y=1/2$ and a single discretization qubit.
Note that for $y < 1$, we need to scale the result by $y$ to adjust for the reduced interval length.

\begin{figure}[ht]
    \centering
    \includegraphics[width=0.49\textwidth]{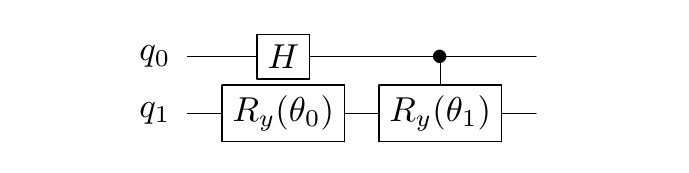}
    \caption{Operator $\mathcal{A}$ for $g(x) = \sin^2(\pi x)$ for $x \in [0, y]$. The angles need to be set according to the chosen discretization as $\theta_0 = 2\pi x_0$ and $\theta_1 = 2 \pi (x_1 - x_0)$.}
    \label{fig:A_for_sin2}
\end{figure}

For the considered problem, the operator $\mathcal{A}$ can be easily implemented using Hadamard gates and (controlled) Pauli Y-rotations $R_y$ as illustrated for one discretization qubit in Fig.~\ref{fig:A_for_sin2}. 
The angles of the (controlled) Y-rotations depend on $x_i$ and are set such that
\begin{eqnarray}
\ket{i}_n\ket{0} \mapsto \ket{i}R_y(2 \pi x_i)\ket{0}.
\end{eqnarray}

\begin{figure}[ht]
    \centering
    \includegraphics[width=0.49\textwidth]{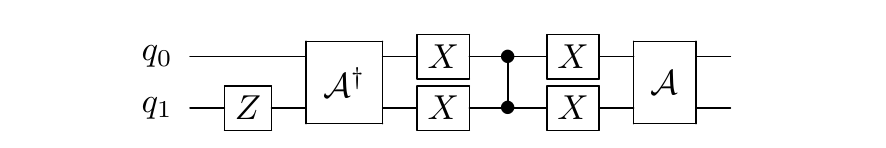}
    \caption{Operator $\mathcal{Q} = \mathcal{A} \mathcal{S}_0 \mathcal{A}^{\dagger} \mathcal{S}_{\psi_0}$ on two qubits, where the second qubit encodes the objective function $g$.}
    \label{fig:Q_for_sin2}
\end{figure}

The reflection $S_{\psi_0}$ can be implemented by a single Pauli Z-gate on the last qubit, while the reflection $S_0$ requires a (multi-)controlled Z-gate preceded and followed by X-gates on all qubits to achieve $\ket{0\ldots0}_{n+1} \mapsto -\ket{0\ldots0}_{n+1}$ instead of $\ket{1\ldots1}_{n+1} \mapsto -\ket{1\ldots1}_{n+1}$.
The resulting $\mathcal{Q}$ operator for one discretization qubit is illustrated in Fig.~\ref{fig:Q_for_sin2}, and it is straightforward to extend this to multiple discretization qubits.

We use simulation as well as a real quantum device, \emph{IBM Q Ourense}, which is accessible through the IBM Quantum Experience via Qiskit, to estimate $\mathbb{P}[\ket{1}]$ for the last qubit of $\mathcal{Q}^k \mathcal{A} \ket{0}_n\ket{0}$, for $n = 1, 2$, and for different values of $k$.
We use MLAE to combine the different measurements to estimate the value of the considered integral.
We set the power of $\mathcal{Q}$ to $k= 2^j$ for $j=0, \ldots, k_{\max}$, for a chosen $k_{\max}$.
The connectivity of IBM Q Ourense as well as the qubits used can be found in Appendix \ref{sec:ibm_q_ourense}, and the corresponding quantum circuits are shown in Appendix \ref{sec:circuits}.

We run every circuit using 8192 shots.
To reduce the noise of the real quantum device, we apply readout error mitigation as well as error mitigation by inserting noisy gates combined with Richardson extrapolation
\cite{Temme2017, Kandala2018, cloudqcomp, Stamatopoulos2019}.
Both techniques are described in more detail in Appendix \ref{sec:error_mitigation}.

\begin{figure}[!htb]
\centering
\includegraphics[width=0.47\textwidth]{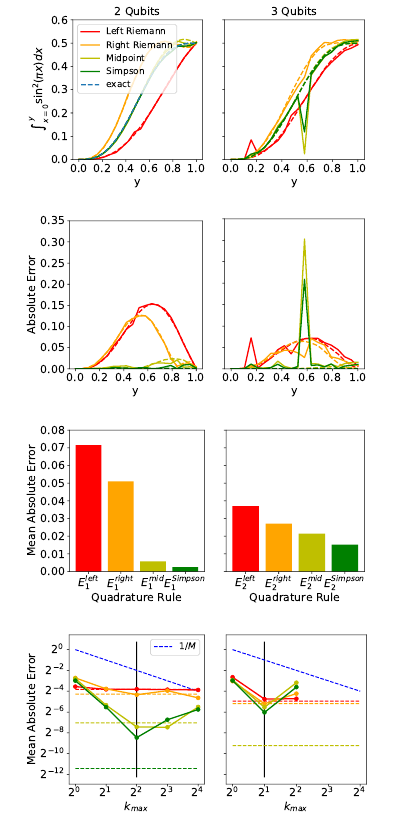}
\caption{
MLAE results for numerical integration.
Left: Results for $n=1$.
Right: Results for $n=2$.
First row: Estimated values for different $y$, using $k_{\max}=2$ (left) and $k_{\max}=1$ (right) - the same parameters were used for rows two and three. 
Second row: Absolute error with respect to the analytic result.
Third row: Mean absolute error over $y$.
Fourth row: Mean absolute error with respect to $k_{\max}$. The blue dashed line indicates $1/M$, i.e., the theoretical convergence rate as reference. The other dashed lines indicate the errors resulting from the analytic evaluation of the quadrature rules. The black vertical line indicates the corresponding choice of $k_{\max}$ for rows one to three.
}
\label{fig:mlae_results}
\end{figure}

We estimate $R_n^{\text{left}}$, $R_n^{\text{right}}$, $R_n^{\text{mid}}$, and combine all three to get $R_n^{\text{Simpson}}$.
The results are shown in Fig.~\ref{fig:mlae_results}, where we compare the analytic solution with the quantum estimates and show aggregated errors to illustrate how the performance changes with an increasing number of qubits and circuit depth.
It can be nicely observed how the approximation error reduces with increasing order of the applied quadrature rule, as well as with the increasing number of discretization points, i.e., qubits.
Furthermore, the quadratic speed-up becomes apparent as we increase $k_{\max}$ until we reach a point where the circuits are getting too long, i.e., the noise distorts the results too much and we do not see further improvements.
Fig.~\ref{fig:mlae_results} also shows another important fact: 
For each quadrature rule there is a threshold, where the model error dominates the QAE estimation error, beyond which it would not make sense to increase $k_{\max}$.
In other words, once that point is reached, the result cannot be improved anymore without increasing the number of discretization points.
For three qubits, the error peaks for a few points. 
The reason for this is the lack of numerical robustness of MLAE, i.e., sometimes a small errors in the measured data can lead to a large deviation of the maximum likelihood estimate.

As mentioned above, our circuits leverage Spin-Echo circuit optimization to reduce the circuit depth.
Tab.~\ref{tab:mlae_circuit_depth} shows the resulting numbers of CNOT gates for two and three qubits, in each case with and without the circuit optimization.
For three qubits analyze all-to-all connectivity, as well as linear connectivity, which is the available layout on IBM Q Ourense for three qubits.
The table shows that the introduced circuit optimization technique can significantly reduce the number of CNOT gates.
Even when compiled to linear connectivity, the resulting number of CNOT gates is not larger than without Spin-Echo circuit optimization for all-to-all connectivity, which equals to the numbers provided in \cite{Suzuki2019}.

\begin{table}[!htb]
    \centering
    \begin{tabular}{cc|cc|ccc}
        \multicolumn{2}{l|}{\#Qubits}  & \multicolumn{2}{c|}{2}  & \multicolumn{3}{c}{3}  \\
        \hline
        \multicolumn{2}{l|}{Topology}  & \hspace{10pt}--\hspace{10pt} & \hspace{10pt}--\hspace{10pt} & all-to-all & linear & all-to-all \\
        \multicolumn{2}{l|}{Optimized} & yes & no & yes & yes & no  \\
        \hline
            &  1 &  4 &  7 &  13 &  17 &  18 \\
            &  2 &  7 & 12 &  23 &  31 &  32 \\
        $k$ &  4 & 13 & 22 &  43 &  59 &  60 \\
            &  8 & 25 & 42 &  83 & 115 & 116 \\
            & 16 & 49 & 82 & 163 & 227 & 228 \\
        
    \end{tabular}
    \caption{Number of CNOT gates to implement $\mathcal{Q}^k \mathcal{A} \ket{0}$: For two qubits, the hardware topology is irrelevant. For three qubits, the hardware topology matters and we compare all-to-all connectivity and linear connectivity (as used on IBM Q Ourense). We show all results with and without Spin-Echo circuit optimization. Note that during the circuit optimization, we can also drop the very last CNOT gate and correct the measurements accordingly with a classical post-processing.}
    \label{tab:mlae_circuit_depth}
\end{table}

For two qubits, i.e., $n=1$, it is interesting to see that we achieve reasonable results for $k_{\max}$ up to 16 (49 CNOT gates) for $R_n^{\text{left}}$, $R_n^{\text{right}}$, and $R_n^{\text{mid}}$.
However, for $R_n^{\text{Simpson}}$, we can only go to $k_{\max}=4$ (13 CNOT gates).
The reason is that the resulting error for the first three methods is not dominated by QAE but by the quadrature rule.
For $R_n^{\text{Simpson}}$ and $k_{\max} > 4$ this seems to change and we cannot improve the results further.
For three qubits, i.e., $n=2$, we get a reasonable behavior for $k_{\max}  \leq 2$ for all four quadrature rules, including $R_n^{\text{Simpson}}$, which corresponds to 31 CNOT gates on the real device.
At a first glance, it might seem contradictory that we can run $R_n^{\text{Simpson}}$ with more CNOT gates on three qubits than on two qubits.
However, as before, the two qubit circuit should result in a much smaller estimation error than the three qubit circuit, which means that the three qubit circuit can tolerate more noise before it starts to dominate the estimated values.

Note that for two qubits, it is possible to optimize the circuit to using only three CNOT gates, independently of $k_{\max}$. This can be achieved by first evaluating the resulting unitary four-by-four matrix and then decomposing it into gates again.
However, this method neither scales well to larger numbers of qubits nor does it provide a benchmark on the number of gates we can apply on the real hardware.
Nevertheless, we performed this experiment as well and report the results in Appendix \ref{sec:hardware_results_2q_compressed}.

\subsection{\label{sec:heston_model} Simulation: Heston Model}

In this section we show how to price a European call option under the \emph{Heston model}.
A European call option gives its owner the right but not the obligation to buy an underlying stock at a fixed time, the maturity $T$, and a fixed price, the strike price $K$.
The Heston model is an example of a stochastic volatility model used in financial mathematics to describe, for instance, the behavior of stock prices \cite{Heston1993}. 
This model was chosen as an illustration, however, the introduced techniques are applicable to any arbitrary stochastic processes as long as the initial and transition probabilities are given as functions that can be (classically) calculated efficiently.

The Heston model consists of a first stochastic process describing the development of the volatility, and a second one -- depending on the volatility -- describing the development of the stock price.
The model is specified by the rate of return $\mu$ of a stock, the long run average price variance $\theta$, the rate $\kappa$ of reverting to $\theta$, the volatility of the volatility $\xi$, and the following two stochastic partial differential equations for the volatility $\nu_t$ and the stock price $S_t$, for $t \in \mathbb{R}_{\geq 0}$:
\begin{eqnarray}
d\nu_t &=& \kappa(\theta - \nu_t)dt + \xi \sqrt{\nu_t}dW_t^{\nu} \\
dS_t &=& \mu S_t dt + \sqrt{\nu_t} S_t dW_t^S,
\end{eqnarray}
where $W_t^S$, $W_t^{\nu}$ are Wiener processes with correlation $\rho$.

Suppose a given initial volatility $\nu_0$ and a given initial stock price $S_0$.
We can then discretize the time $t = 0, 1, 2, \ldots$ with a time step size $\delta t$ and we derive the discrete transition laws as
\begin{eqnarray}
\nu_{t+1} &=& \nu_t + \kappa(\theta - \nu_t)\delta t + \xi \sqrt{\nu_t}X_t^{\nu} \\
S_{t+1}   &=& S_t   + \mu S_t \delta t + \sqrt{\nu_t}S_t X_t^S,
\end{eqnarray}
where
\begin{eqnarray}
(X_t^S, X_t^{\nu}) &\sim& \mathcal{N}
\left(
\left(
\begin{array}{c} 0 \\ 0 \end{array}
\right),
\delta t\left(
\begin{array}{cc}
1 & \rho \\
\rho & 1
\end{array}
\right)
\right).
\end{eqnarray}
This allows us to derive the conditional probabilities
\begin{eqnarray}
f_t\left(\nu_t, S_t \mid \nu_{t-1}, S_{t-1}\right)
\end{eqnarray}
as the PDFs of two-dimensional normal distributions.

Given a model for the stock price at maturity of the option, i.e., for $S_T$, the expected payoff of a European call option is given by
\begin{eqnarray}
\mathbb{E}[\max\{S_T - K, 0\}],
\end{eqnarray}
and equals the fair option price before discounting.
For a constant interest rate, discounting just results in a simple correction term to compute the present value of the future payoff.
It is also straightforward to extend to stochastic interest rates. We can add another stochastic process for the discount rate and then use quantum arithmetic to discount the expected payoff and directly use QAE to estimate the fair option price.
However, for ease of presentation, we ignore discounting and just focus on the expected payoff.

For an illustrative example, we now set $K = 1$, $\delta t = 1$, and $T = 2$, i.e., $t = 0, 1, 2$.
Furthermore, we assume $\nu_0 = 1, S_0 = 1$ and set 
$\kappa = 1$,
$\theta = 1$,
$\xi = 0.5$,
$\mu = 1$, and $\rho = 0$.
We discretize $\nu_1 \in \{ 0.8, 1.2 \}$ and $S_1 \in \{ 0.75, 1.25 \}$ using one qubit each, and $S_2 \in \{ 0, 1, 2, 3 \}$ using two qubits.
Note that we do not need to represent $\nu_2$ since $S_2$ does not depend on it, and thus, it does not contribute to the option price.
Thus, we use four qubits in total to discretize the volatility and stock price processes.

Given the grid points $x_{i-1}$, $x_i$, $x_{i+1}$ for some index $i$, and the continuous probability density functions $f_t$ introduced above, the probability for $x_i$ is defined as the probability of the interval $[(x_{i-1}+x_i)/2, (x_i + x_{i+1})/2]$.
In case $x_i$ is the first (last) grid point, the lower bound (upper bound) is replaced by negative (positive) infinity.

\begin{figure}[ht]
\centering
\includegraphics[width=0.49\textwidth]{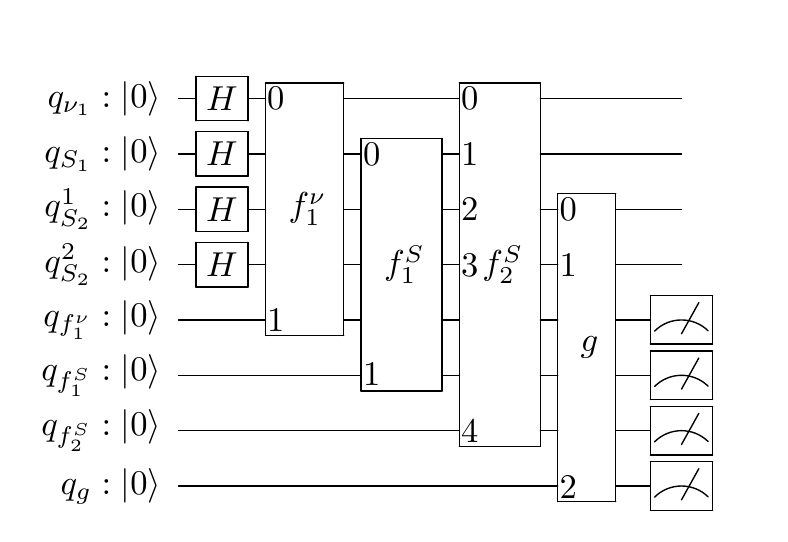}
\caption{$\mathcal{A}$-circuit for a European call option under the Heston model: First, Hadamard gates are applied to the state qubits to span the equal superposition for the state discretization. Then, we apply the rotations corresponding to the different probability density functions, where the last qubit of the gate always denotes the target. Last, we apply a gate corresponding to the objective function $g(S_2) = \max\{S_2 - K, 0\}$. The four measurements indicate that we need to take into account the last four qubits to estimate the expected payoff. More precisely, the probability of measuring $\ket{1111}$ corresponds to the (normalized) expected payoff we are interested in.}
\label{fig:heston_model_european_call_circuit}
\end{figure}

For given $\nu_{t-1}$ and $S_{t-1}$, and assuming $\rho = 0$, $\nu_t$ and $S_t$ are independent.
Thus, we can split the functions $f_t$ into $f_t^{\nu}$ and $f_t^{S}$ to simplify the computation and we leverage the multiplication technique introduced in Sec.~\ref{sec:state_preparation}.
We add three ancilla qubits to represent $f_1^{\nu}$, $f_1^{S}$, $f_2^{S}$ and another one to represent the actual payoff, i.e., four ancillas in total.
The resulting probabilities are provided as reference in Tab.~\ref{tab:heston_probabilities}.
Note that we need to normalize the payoff function $g(S_2) = \max\{S_2 - K, 0\}$ such that it takes values in $[0, 1]$. Given $K = 1$ and the range for $S_3$, this implies that we need to divide $g$ by two.
The resulting eight-qubit circuit is illustrated in Fig.~\ref{fig:heston_model_european_call_circuit}.
For simplicity, we use uniformly controlled Pauli rotations as provided by Qiskit \cite{iten2019, Qiskit} to implement the different operations, i.e., we pre-compute the rotation angles for the different cases. 
In general, we would need to compute the actual transition probabilities using quantum arithmetic.

\begin{table}[ht]
    \centering
    \begin{tabular}{cc|ccccc}
        && \multicolumn{5}{c}{$\mathbb{P}[S_2 = x \mid \nu_1, S_1]$} \\
        \hline
        $\nu_1$ ($\mathbb{P}$) & $S_1$ ($\mathbb{P}$) & $x=$ & 0 & 1 & 2 & 3  \\
        \hline
        0.8 (0.50) & 0.75 (0.38) && 0.063 & 0.937 & 0.001 & 0.000 \\ 
                   & 1.25 (0.62) && 0.007 & 0.631 & 0.361 & 0.001 \\
        1.2 (0.50) & 0.75 (0.38) && 0.105 & 0.890 & 0.005 & 0.000 \\
                   & 1.25 (0.62) && 0.022 & 0.592 & 0.382 & 0.005 \\
        \hline
        \multicolumn{2}{c|}{marginal distribution} && 0.040 & 0.725 & 0.233 & 0.002
    \end{tabular}
    \caption{Probabilities. This table provides the probabilities of certain values for $\nu_1$ and $S_1$, which are independent variables, as well as the conditional probabilities for $S_2$ given $\nu_1$ and $S_1$.}
    \label{tab:heston_probabilities}
\end{table}

We simulate the circuit introduced above by using the simulators provided by Qiskit, and evaluate the probability of measuring $\ket{1111}$ in the last four qubits, which equals $0.1185$. This can be easily verified using the marginal distribution for $S_2$ provided in Tab.~\ref{tab:heston_probabilities} and the (normalized) payoff function $g(S_2)$.
Thus, the operator $\mathcal{A}$ illustrated in Fig.~\ref{fig:heston_model_european_call_circuit} corresponds to pricing a European call option under the Heston model and can directly be used with QAE and its variants to achieve a quadratic speed-up over classical Monte Carlo simulation.

Within this section, we focused on a European call option, i.e., a very simple type of option.
However, it is straightforward to extend the results in this paper to more exotic options, e.g., path-dependent options, following the techniques introduced in \cite{Stamatopoulos2019}.
Similarly, the Heston model was only used as illustration and we can extend the methodology to arbitrary stochastic processes.

\section{\label{sec:conclusions} Conclusions}

In the present paper we introduced an efficient approach to preparing quantum states for QAE and showed how basic numerical integration can help to reduce the approximation error while keeping the number of qubits constant.
Furthermore, we developed a generic circuit optimization technique for QAE and demonstrated our insights on a simple numerical integration problem using real quantum hardware as well as on a relevant model from financial mathematics using simulation.

This is a significant enhancement of the state-of-the-art. We do not require the probability distribution functions to be log-concave and do not impose any other requirements on their structure except being efficiently computable.
Note that our approach is also more efficient than the loading scheme for log-concave functions proposed in \cite{Grover2002}.
The improvement we demonstrate in this paper is only possible because we are not treating state preparation separately, but in the context of QAE, i.e., together with the algorithm where the prepared state is being used.

Determining the most efficient loading scheme will depend on the exact situation and is a task for future research.
It is likely that this will result in a combination of different approaches automatically constructed by future quantum compilers.

\begin{acknowledgments}
We would like to thank David Sutter and Dmitri Maslov for the constructive technical discussions on data loading and circuit optimization. 
We further acknowledge the support of the National Centre of Competence in Research \textit{Quantum Science and Technology} (QSIT).

IBM, IBM Q, and Qiskit are trademarks of International Business Machines Corporation, registered in many jurisdictions worldwide. Other product or service names may be trademarks or service marks of IBM or other companies.
\end{acknowledgments}




\appendix

\section{IBM Q Ourense}\label{sec:ibm_q_ourense}

\begin{figure}[ht]
\centering
\includegraphics[width=0.25\textwidth]{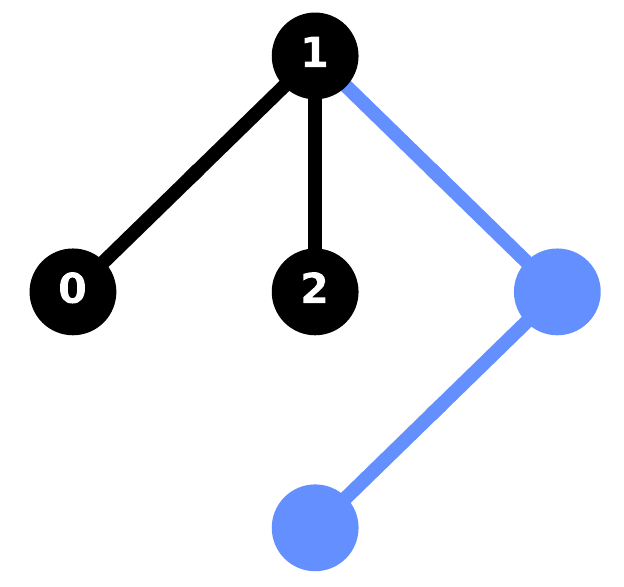}
\caption{
The connectivity of IBM Q Ourense. For the two-qubit experiments we used qubits $0$ and $1$, and for the three-qubit experiments we used qubits $0$, $1$ and $2$.
}
\label{fig:ourense_connectivity}
\end{figure}

Fig.~\ref{fig:ourense_connectivity} shows the connectivity of IBM Q Ourense, the quantum device used for the experiments on real hardware, as well as the qubits used.

\section{Quantum Circuits}\label{sec:circuits}

\begin{figure*}[!htb]
\centering
\includegraphics[width=0.6\textwidth]{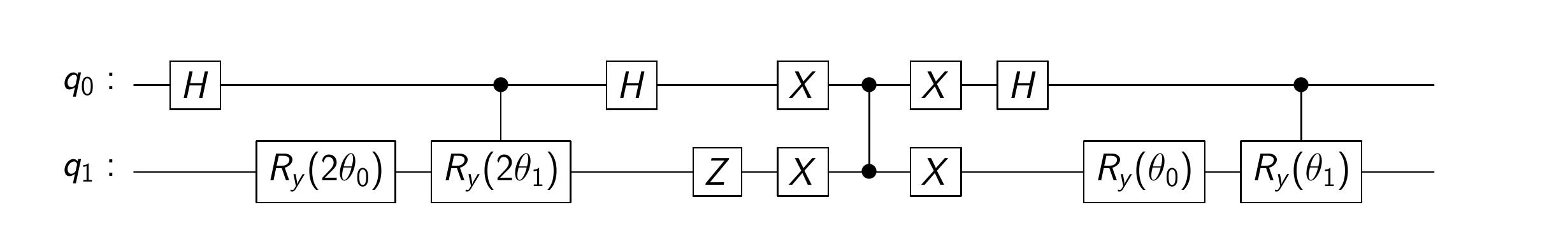}
\caption{
Spin-Echo optimized circuit for $\mathcal{Q}\mathcal{A}\ket{0}$ for two qubits. The angles need to be set, according to the discretization, as $\theta_0 = 2\pi x_0$ and $\theta_1 = 2\pi (x_1 - x_0)$. Note that the first set of $R_y$ rotations has a factor of $2$ for the angle due to the Spin-Echo optimization.
}
\label{fig:circuit_2q_QA}
\end{figure*}

\begin{figure*}[!htb]
\centering
\includegraphics[width=0.8\textwidth]{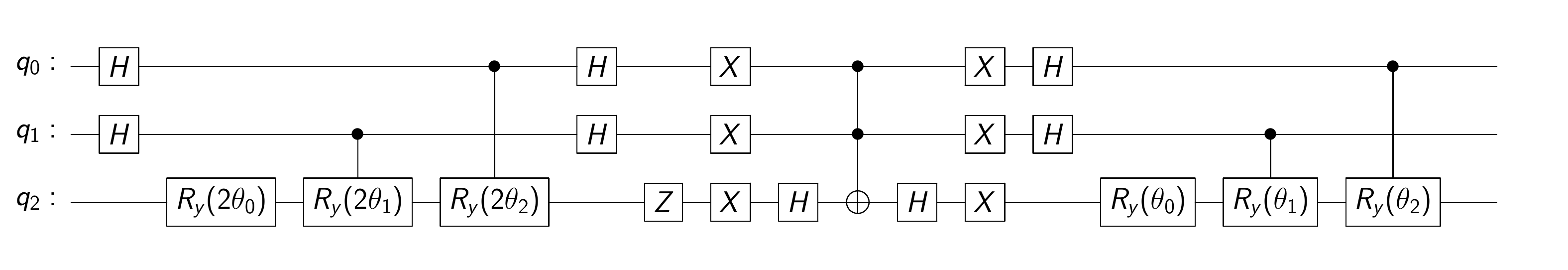}
\caption{
Spin-Echo optimized circuit for $\mathcal{Q}\mathcal{A}\ket{0}$ for three qubits. The angles need to be set, according to the (equidistant) discretization, as $\theta_0 = 2\pi x_0$, $\theta_1 = 2\pi (x_1 - x_0)$ and $\theta_2 = 2\pi (x_2 - x_0)$. Note that the first set of $R_y$ rotations has a factor of $2$ for the angle due to the Spin-Echo optimization.
}
\label{fig:circuit_3q_QA}
\end{figure*}

This section describes the quantum circuits that were evaluated in Sec.~\ref{sec:hardware_results}.
Fig.~\ref{fig:circuit_2q_QA} and Fig.~\ref{fig:circuit_3q_QA} show the circuit for $\mathcal{Q}\mathcal{A}\ket{0}$, for two and three qubits, respectively.

It is straightforward to expand to multiple applications of $\mathcal{Q}$ by repeating $\mathcal{Q}$ and taking into account the Spin-Echo circuit optimization as indicated in the figures.

\section{Error Mitigation}\label{sec:error_mitigation}

We apply the same strategy to mitigate errors as, for instance, also used in \cite{Stamatopoulos2019}.
First, readout errors are mitigated by running a calibration sequence over all basis states to construct a matrix representing the conditional distribution of measurements given a prepared state.
This can subsequently be used to correct the measurements in our experiments.
More details can be found in \cite{Qiskit, dewes}.
Second, to mitigate the error of CNOT gates, we first amplify the noise and then extrapolate to the zero noise limit.
More precisely, we run a circuit as given, then we replace every CNOT gate by three CNOT gates, and last, by five CNOT gates.
In theory, inserting these gates should have no effect, since an odd number of CNOT gates should be equal to a single CNOT gate.
In practice, this amplifies the error of the CNOT gates to three times and five times the original error.
Having these three data points with the increasing error allows us to do a quadratic extrapolation to the zero-noise limit, which leads to the results shown in this paper.
More details on this technique can be found in \cite{Temme2017, Kandala2018, cloudqcomp}.

\section{Quantum Hardware Results for 2 Qubits with 3 CNOT Gates}\label{sec:hardware_results_2q_compressed}

\begin{figure*}[!htb]
\centering
\includegraphics[width=0.6\textwidth]{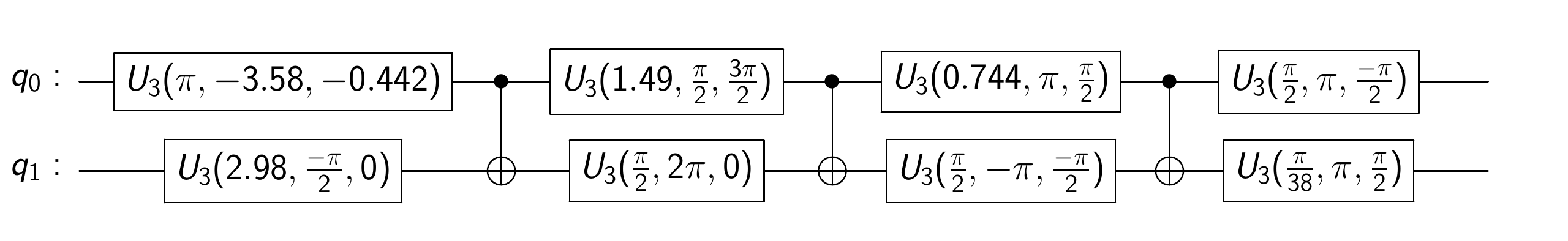}
\caption{
The numerical integration circuit for the mid-point rule with $k=1$ using one qubit for discretization after optimizing to only three CNOT gates.
}
\label{fig:2q_compressed_circuit}
\end{figure*}

This section shows the results for the numerical integration problem introduced in Sec.~\ref{sec:hardware_results} for the 2-qubit circuit optimized to use only three CNOT gates in total.
This can be achieved by first classically evaluating the corresponding unitary matrix and then decomposing it into quantum gates again, for instance using the functionality provided by Qiskit. 
The resulting circuit for $k=1$ is shown in Fig.~\ref{fig:2q_compressed_circuit}.
It should be noted that this approach does not scale to larger numbers of qubits, and is reported here for comparison only.

\begin{figure}
\centering
\includegraphics[width=0.44\textwidth]{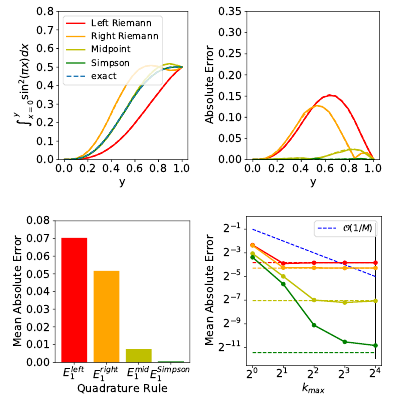}
\caption{
MLAE results for 2-qubits compressed to 3 CNOT gates:
Top left: Estimated values for different $y$, using $k_{\max}=16$ (same for top right / bottom left).
Top right: Absolute error with respect to analytic result.
Bottom left: Mean absolute error over $y$.
Bottom right: Mean absolute errors with respect to $k_{\max}$. The blue dashed line indicates $\mathcal{O}(1/M)$. The other dash lines indicate errors resulting from the analytic evaluation of the quadrature rules. The black vertical line indicates the corresponding choice of $k_{\max}$ for the other figures.
}
\label{fig:mlae_results_2q_compressed}
\end{figure}

As before, the circuits were run on IBM Q Ourense using $8192$ shots and error mitigation via insertion of noisy CNOT gates (cf.~Sec.~\ref{sec:hardware_results} and Appendix \ref{sec:error_mitigation}).
Fig.~\ref{fig:mlae_results_2q_compressed} shows the results in the same way as presented in Sec.~\ref{sec:hardware_results}.
It can be seen that we get good results for all $k_{\max}$, although the convergence of Simpson's rule starts to slow down towards $k_{\max}=16$, which is likely due to the remaining errors of the corresponding circuit starting to dominate as the estimated result gets more and more accurate.
Nevertheless, we can estimate the integral with an average absolute error smaller than $2^{-10} \approx 10^{-3}$ using Simpson's rule.

\bibliography{bibliography}

\end{document}